\documentclass[prd,twocolumn,nofootinbib]{revtex4}

\usepackage{amsmath,amssymb}
\usepackage{mathrsfs}
\usepackage{mathtools}
\usepackage{rotating}
\usepackage{mathrsfs}
\usepackage{pstricks}
\usepackage{pst-func}
\usepackage{graphicx}

\def\BibTeX{{\rm B\kern-.05em{\sc i\kern-.025em b}\kern-.08em
    T\kern-.1667em\lower.7ex\hbox{E}\kern-.125emX}}

\begin{document}

\title{Reply to ``Comment on `Dr.~Bertlmann's Socks in a Quaternionic World of Ambidextral Reality'''}

\author{Joy Christian}

\email{jjc@bu.edu}

\affiliation{Einstein Centre for Local-Realistic Physics, 15 Thackley End, Oxford OX2 6LB, United Kingdom}

\begin{abstract}
In this paper, I respond to a critique of one of my papers previously published in this journal, entitled `Dr.~Bertlmann's socks in a quaternionic world of ambidextral reality.' The geometrical framework presented in my paper is based on a quaternionic 3-sphere, or $S^3$, taken as a model of the physical space in which we are inescapably confined to perform all our experiments. The framework intrinsically circumvents Bell's theorem by reproducing the singlet correlations local-realistically, without resorting to backward causation, superdeterminism, or any other conspiracy loophole. In this response, I demonstrate point by point that, contrary to its claims, the critique has not found any mistakes in my paper, either in the analytical model of the singlet correlations or in its event-by-event numerical simulation based on Geometric Algebra. 
\end{abstract}

\maketitle

\renewcommand\bf{\mathbf}

\section{Introduction}\label{intro}

Mathematical constructions are often obfuscated rather than enlightened by their misplaced criticisms, and the recent critique in \cite{Gill-Ieee} of the geometrical 3-sphere framework for local-realistically underpinning quantum correlations I have presented in \cite{Disproof,IJTP,RSOS,IEEE-1,IEEE-2} is no exception. Therefore, in Section~\ref{I} below I first summarize the 3-sphere framework in some detail before responding in Section~\ref{II} to the critique in \cite{Gill-Ieee}. While the critique is focused on the 3-sphere model presented in my latest paper \cite{IEEE-2}, it does not raise any new questions. The issues raised therein have been extensively addressed by me already, for example in \cite{Reply-Gill,Reply-Gill-IJTP,Scott}. It is unfortunate that the critique in \cite{Gill-Ieee} does not directly address any of my previous responses. In this regard, my response \cite{Reply-Lasenby} to a separate critique is also of interest. In addition to these responses, the critique in \cite{Gill-Ieee} has also overlooked the large number of specific questions that are answered in the appendices of  \cite{IEEE-1} and \cite{IEEE-2}.

The motivations behind the critique in \cite{Gill-Ieee} seem to stem from the fact that the geometrical framework I have presented in \cite{Disproof,IJTP,RSOS,IEEE-1,IEEE-2} appears to contradict the well known mathematical theorem proposed by Bell \cite{Bell-1964}. But as we shall soon see, the facts are not as black and white as the critique has claimed. The proof of Bell's theorem relies on the strong correlations predicted by the entangled quantum states, such as the singlet state. The claim of the theorem is that no local and realistic theory of the kind espoused by Einstein can reproduce all of the statistical predictions of quantum mechanics, especially those that are predicted using quantum entanglement. But in \cite{Disproof,IJTP,RSOS,IEEE-1,IEEE-2}, I have demonstrated that the strong correlations observed in Nature, including those predicted by the singlet state, have little to do with quantum entanglement {\it per se}. In general, they are local and realistic correlations among the limiting scalar points of an octonion-like 7-sphere \cite{RSOS,Eight}, which is an algebraic representation space of a quaternionic 3-sphere, taken as the physical space. Thus, the assumptions underlying \cite{Disproof,IJTP,RSOS,IEEE-1,IEEE-2} and those of Bell's theorem are different.

By contrast, in its analysis the critique in \cite{Gill-Ieee} begins with an incorrect version of the above model by writing some of its equations incorrectly within a flat Euclidean space ${\mathrm{I\!R}^3}$. It then derives a constant value of the singlet correlations, ${\cal E}({\bf a},\,{\bf b})=-1$ for all detector directions ${\bf a}$ and ${\bf b}$, by failing to respect the geometrical properties of the 3-sphere, such as the spinorial sign changes in the quaternions, criticizes this incorrect value, and concludes that it has thereby criticized the 3-sphere model of \cite{Disproof,IJTP,RSOS,IEEE-1,IEEE-2}. In the process, the critique also violates the conservation of zero spin angular momentum. 

But before I bring out such oversights from the critique \cite{Gill-Ieee}, in the next section I first review the status of Bell's theorem in order to highlight in which sense it is circumvented by the 3-sphere model of quantum correlations presented in \cite{Disproof,IJTP,RSOS,IEEE-1,IEEE-2}.

\section{Boole's inequality versus Bell's theorem}\label{Boole}

Much is made in the critique \cite{Gill-Ieee} of the so-called ``theorem'' of Bell that claims that the models such as the one presented in \cite{Disproof,IJTP,RSOS,IEEE-1,IEEE-2} are impossible. But mathematically a theorem with loopholes \cite{loopholes} is an oxymoron, while physically we know that the bounds on Bell inequalities are not respected by Nature. The consequent conclusion that therefore Nature must be non-local, non-realistic, or conspiratorial is not justified. For Bell's theorem depends on a number of assumptions \cite{RSOS}, in addition to those of locality and realism. And, in fact, Bell inequalities can be derived without assuming either locality or realism, as shown, for example, in Section 4.2 of \cite{RSOS}.

That is not to say that Bell's theorem \cite{Bell-1964} does not have a sound mathematical core. When stated as a mathematical theorem in probability theory, there can be no doubt about its validity. 
But my work on the subject \cite{Disproof,IJTP,RSOS,IEEE-1,IEEE-2} does not challenge this mathematical core, if it is viewed as a piece of mathematics. What it challenges are the metaphysical conclusions regarding locality and realism derived from that mathematical core. My work thus draws a sharp distinction between the mathematical core of Bell's theorem and the metaphysical conclusions derived from it. Let me unpack these remarks to explain in what sense the local-realistic framework presented in \cite{Disproof,IJTP,RSOS,IEEE-1,IEEE-2}
circumvents Bell's theorem.

As acknowledged in the critique in \cite{Gill-Ieee}, the mathematical core of Bell's theorem goes back to Boole's derivation of an inequality within probability theory, one hundred and eleven years before the publication of Bell's theorem \cite{Boole-1,Boole-2}. In the modern form, it is the famous Bell-CHSH \cite{Clauser} inequality 
\begin{equation}
-2\,\leqslant\,{\cal E}({\bf a},\,{\bf b})\,+\,{\cal E}({\bf a},\,{\bf b'})\,+\,{\cal E}({\bf a'},\,{\bf b})\,-\,{\cal E}({\bf a'},\,{\bf b'})\,\leqslant\,+2\,, \label{CHSH}
\end{equation}
where each of the expectation values is defined as the average 
\begin{equation}
{\cal E}({\bf a},\,{\bf b})\,=\int_{\Lambda}
{\mathscr A}({\bf a},\,\lambda)\,{\mathscr B}({\bf b},\,\lambda)\;\,d\rho(\lambda) \label{prob} 
\end{equation}
that satisfies the EPR's condition of perfect anti-correlation:
\begin{equation}
{\cal E}({\bf n},\,{\bf n}) = -1, \;\;\;\forall\;{\bf n}\in S^2\subset{\mathrm{I\!R}^3}. 
\end{equation}
Here $\lambda\in\Lambda$ denotes a complete specification of the physical state of the singlet system at a suitable instant, ${\rho(\lambda)}$ denotes the normalized probability measure on the space ${\Lambda}$ of the complete states, and ${\mathscr A({\mathbf{a}},\,{\lambda})=\pm1}$ and ${\mathscr B({\mathbf{b}},\,{\lambda})=\pm1}$ are the measurement functions specifying the results observed by Alice and Bob for a given run of the experiment. These functions satisfy the following conditions of local causality:
\begin{quote}
\underbar{Local Causality}: Apart from the initial state or a hidden variable ${\lambda}$, the result ${{\mathscr A}=\pm1}$ of Alice depends {\it only} on the measurement direction ${\bf a}$, chosen freely by Alice, regardless of Bob's actions. And similarly, apart from the initial state ${\lambda}$, the result ${{\mathscr B}=\pm1}$ of Bob depends {\it only} on the measurement direction ${\bf b}$, chosen freely by Bob, regardless of Alice's actions. In particular, the function ${{\mathscr A}({\bf a},\,\lambda)}$ {\it does not} depend on ${\bf b}$ or ${\mathscr B}$, the function ${{\mathscr B}({\bf b},\,\lambda)}$ {\it does not} depend on ${\bf a}$ or ${\mathscr A}$, and, moreover, the initial state ${\lambda}$ does not depend on ${\bf a}$, ${\bf b}$, ${\mathscr A}$, or ${\mathscr B}$.
\end{quote}
In addition, the probability measure ${\rho(\lambda)}$ in (\ref{prob}) is required to depend {\it only} on ${\lambda}$, and not on either ${\bf a}$ or ${\bf b}$, which are the freely chosen detector settings at the time of measurements.

Although I have introduced quite a bit of physics in the foregoing, the mathematical core of Bell's theorem is rather simple. It can be stated simply as the claim that, for the product ${\mathscr A({\mathbf{a}},\,{\lambda}){\mathscr B({\mathbf{b}},\,{\lambda})}=\pm1}$ of any local-realistic functions ${\mathscr A({\mathbf{a}},\,{\lambda})=\pm1}$ and ${\mathscr B({\mathbf{b}},\,{\lambda})=\pm1}$, the expectation value
\begin{align}
{\cal E}({\mathbf{a}},\,{\mathbf{b}})\,&=\lim_{\,n\,\gg\,1}\left[\frac{1}{n}\sum_{k\,=\,1}^{n}\,
{\mathscr A}({\mathbf{a}},\,{\lambda}^k)\;{\mathscr B}({\mathbf{b}},\,{\lambda}^k)\right] \notag \\
&\equiv\,\Bigl\langle\,{\mathscr A}_{k}({\mathbf{a}})\,{\mathscr B}_{k}({\mathbf{b}})\,\Bigr\rangle \notag \\
&= -{\bf a}\cdot{\bf b}
\end{align}
is impossible to achieve, because it would lead to ``violations'' of the bounds of $\pm2$ claimed in (\ref{CHSH}) on the CHSH correlator. The expectation values are therefore constrained to be within
\begin{align}
{\cal E}({\mathbf{a}},\,{\mathbf{b}})=\,
\begin{cases}
-\,1\,+\,\frac{2}{\pi}\,\eta_{{\bf a}{\bf b}}
\;\;\;\text{if} &\!\! 0 \leqslant \eta_{{\bf a}{\bf b}} \leqslant \pi \\
\\
+\,3\,-\,\frac{2}{\pi}\,\eta_{{\bf a}{\bf b}}
\;\;\;\text{if} &\!\! \pi \leqslant \eta_{{\bf a}{\bf b}} \leqslant 2\pi\,, \label{equawhichr-2}
\end{cases}
\end{align}
where $\eta_{{\bf a}{\bf b}}$ is the angle between the detector directions ${\bf a}$ and ${\bf b}$. Here I have rewritten the integral in (\ref{prob}) as a discrete sum because that is what is observed in the experiments, with ${\lambda^k}$ being an initial state for the ${k^{\mathrm{th}}}$ run of the experiment.

Now, the proof of Bell's claim follows from {\it the additivity of expectation values}, which allows us to equate the sum of four {\it separate} averages of numbers $+1$ and $-1$ appearing in the CHSH inequality (\ref{CHSH}) with a single average of their sum:
\begin{align}
{\cal E}({\mathbf{a}},\,&{\mathbf{b}})\,+\,{\cal E}({\mathbf{a}},\,{\mathbf{b'}})\,+\,{\cal E}({\mathbf{a'}},\,{\mathbf{b}})\,-\,{\cal E}({\mathbf{a'}},\,{\mathbf{b'}}) \notag \\
&=\Bigl\langle\,{\mathscr A}_{k}({\mathbf{a}})\,{\mathscr B}_{k}({\mathbf{b}})\,\Bigr\rangle + \Bigl\langle\,{\mathscr A}_{k}({\mathbf{a}})\,{\mathscr B}_{k}({\mathbf{b'}})\,\Bigr\rangle \notag \\
&\;\;\,\;\;\;\;\;\;+\Bigl\langle\,{\mathscr A}_{k}({\mathbf{a'}})\,{\mathscr B}_{k}({\mathbf{b}})\,\Bigr\rangle - \Bigl\langle\,{\mathscr A}_{k}({\mathbf{a'}})\,{\mathscr B}_{k}({\mathbf{b'}})\,\Bigr\rangle \label{four} \\
&=\Bigl\langle{\mathscr A}_{k}({\mathbf{a}}){\mathscr B}_{k}({\mathbf{b}})+{\mathscr A}_{k}({\mathbf{a}}){\mathscr B}_{k}({\mathbf{b'}}) \notag \\
&\;\;\;\;\;\;\;\;\;\;\;\;\;\;\;\;\;\;+{\mathscr A}_{k}({\mathbf{a'}}){\mathscr B}_{k}({\mathbf{b}})-{\mathscr A}_{k}({\mathbf{a'}}){\mathscr B}_{k}({\mathbf{b'}})\Bigr\rangle. \label{reppp}
\end{align}
This immediately reduces the sum (\ref{four}) of four averages to
\begin{equation}
\Bigl\langle{\mathscr A}_{k}({\mathbf{a}})\big\{{\mathscr B}_{k}({\mathbf{b}})+{\mathscr B}_{k}({\mathbf{b'}})\big\}\,+\,{\mathscr A}_{k}({\mathbf{a'}})\big\{{\mathscr B}_{k}({\mathbf{b}})-{\mathscr B}_{k}({\mathbf{b'}})\big\}\Bigr\rangle.\label{absurd}
\end{equation}
And because ${{\mathscr B}_{k}({\mathbf{b}})=\pm1}$, if ${|{\mathscr B}_{k}({\mathbf{b}})+{\mathscr B}_{k}({\mathbf{b'}})|=2}$, then ${|{\mathscr B}_{k}({\mathbf{b}})-{\mathscr B}_{k}({\mathbf{b'}})|=0}$, and vice versa. Consequently, using ${{\mathscr A}_{k}({\mathbf{a}})=\pm1}$, it is easy to see that the absolute value of the above average cannot exceed 2, just as Bell concluded \cite{Clauser}: 
\begin{align}
-\,2\,\leqslant\,\Bigl\langle\,&{\mathscr A}_{k}({\mathbf{a}})\,{\mathscr B}_{k}({\mathbf{b}})\,+\,{\mathscr A}_{k}({\mathbf{a}})\,{\mathscr B}_{k}({\mathbf{b'}}) \notag \\
&+\,{\mathscr A}_{k}({\mathbf{a'}})\,{\mathscr B}_{k}({\mathbf{b}})\,-\,{\mathscr A}_{k}({\mathbf{a'}})\,{\mathscr B}_{k}({\mathbf{b'}})\,\Bigr\rangle\,\leqslant\,+\,2\,.\label{5-1}
\end{align}
On the other hand, if we substitute ${\cal E}({\mathbf{a}},\,{\mathbf{b}})=-\,{\mathbf{a}}\cdot{\mathbf{b}}$, {\it etc.}, into (\ref{four}), then it is easy to demonstrate that the bounds of $\pm2$ in (\ref{CHSH}) can be exceeded for some detector directions, giving
\begin{equation}
-2\sqrt{2}\leqslant{\cal E}({\mathbf{a}},\,{\mathbf{b}})+{\cal E}({\mathbf{a}},\,{\mathbf{b'}})+{\cal E}({\mathbf{a'}},\,{\mathbf{b}})-{\cal E}({\mathbf{a'}},\,{\mathbf{b'}})\leqslant 2\sqrt{2}. \label{strong}
\end{equation}
Consequently, according to Bell's theorem, the quantum mechanical correlations ${\cal E}({\mathbf{a}},\,{\mathbf{b}})=-\,{\mathbf{a}}\cdot{\mathbf{b}}$ are impossible to achieve within the local-realistic framework specified above. 

The above proof, however, while mathematically sound, harbors a profound physical mistake. The mistake is hidden in the assumption (\ref{reppp}) of the additivity of expectation values, and it is the same mistake that von Neumann's ex-theorem against the general hidden variable theories was based on, as I have explained in \cite{Oversight}. Both von Neumann's theorem and Bell's theorem unjustifiably assume the additivity of expectation values within hidden variable theories to derive their respective conclusions. However, for observables that are not simultaneously measurable, such as those involved at one of the stations in the Bell-test experiments, the equivalence of the sum of expectation values and the expectation value of the sum of corresponding measurement results, although respected within quantum mechanics, need not hold within hidden variable theories, as noted by Einstein \cite{Oversight}. Once this oversight is removed from Bell's argument and local realism is implemented correctly, the bounds on the CHSH correlator derived in Eq.~(\ref{5-1}) above work out to be $\pm2\sqrt{2}$ instead of $\pm2$, thereby mitigating the conclusion of Bell’s theorem \cite{Oversight}. Consequently, what is ruled out by Bell-test experiments is not local realism but the additivity (\ref{reppp}) of expectation values. 

\section{Review of the $S^3$ Model of
correlations}\label{I}

Let us now review the quaternionic 3-sphere model for the strong correlations proposed in \cite{Disproof,IJTP,RSOS,IEEE-1,IEEE-2}. As far as the correlations predicted by the singlet state are concerned (which are the focus of \cite{Gill-Ieee} and \cite{IEEE-2}), it will be sufficient to restrict to the 3-sphere instead of its algebraic representation space $S^7$  considered in \cite{RSOS}. A quaternionic 3-sphere can be defined as
\begin{equation}
S^3:=\left\{\,{\bf q}(\psi,\,{\bf r}):=\exp\left[\,{\bf J}({\bf r})\,\frac{\psi}{2}\,\right]
\Bigg|\;||\,{\bf q}(\psi,\,{\bf r})\,||^2=1\right\}\!, \label{nonsin}
\end{equation}
where ${{\bf J}({\bf r})}$ is a bivector (or pure quaternion) rotating about ${{\bf r}\in{\mathrm{I\!R}}^3}$ with the rotation angle ${\psi}$ in the range ${0\leq\psi < 4\pi}$. Here the notations and conventions of Geometric Algebra are used \cite{Clifford}. Now, the central hypothesis I have put forward in \cite{IEEE-1} and \cite{IEEE-2} is that the strong correlations we observe in the Bell-test experiments are consequences of the fact that three-dimensional physical space is best modeled as a closed and compact quaternionic 3-sphere, $S^3$, using Geometric Algebra, as in (\ref{nonsin}), rather than as a flat and open space ${\mathrm{I\!R}^3}$ using ``vector algebra.'' This is by no means an {\it ad hoc} hypothesis. Note that ${S^3}$ happens to be isomorphic to the spatial part of one of the well known cosmological solutions of Einstein's field equations of general relativity, representing a closed universe with positive curvature \cite{IEEE-1,IEEE-2}. It is universally accepted that the spacetime geometries of our universe are described by the Friedmann-Robertson-Walker line element
\begin{equation}
ds^2=dt^2-a^2(t)\,d\Sigma^2, \;\;\,d\Sigma^2=\left[\frac{d\rho^2}{1-\kappa\,\rho^2}+\rho^2 d\Omega^2\right]\!, \label{frw}
\end{equation}
where ${a(t)}$ is the scale factor, ${\Sigma}$ is a spacelike hypersurface, ${\rho}$ is the radial coordinate within ${\Sigma}$, ${\kappa}$ is the ``normalized'' curvature of ${\Sigma}$, and ${\Omega}$ is a solid angle within ${\Sigma}$ \cite{d'Inverno}. Since we are primarily concerned with a galactic, solar, or terrestrial scenario, in what follows, without loss of generality, we will restrict our attention to the current epoch of the cosmos by setting the scale factor ${a(t) = 1}$ in the above line element. It then allows three possible geometries for the spacetime with the product topology ${\mathrm{I\!R}\times\Sigma}$ so that the corresponding spacelike hypersurfaces $\Sigma$ can be isomorphic only to ${\mathrm{I\!R}^3}$, $S^3$, or $H^3$, with $H^3$ being a hyperboloid of negative curvature. Among these possible three-geometries, only $S^3$ represents a closed universe with compact geometry and constant positive curvature. Moreover, the cosmic microwave background spectra  recently  mapped  by  the space observatory, {\it Planck}, now prefers a positive curvature at more than  99\%  confidence  level \cite{closed,Handley}. And yet, topologically $S^3$ can be constructed by adding only a single mathematical point to ${{\mathrm{I\!R}}^3}$ at infinity:
\begin{equation}
S^3 = \,{\mathrm {I\!R}}^3 \cup \left\{\infty\right\}.
\end{equation}
What is more, unlike ${\mathrm{I\!R}^3}$ and $H^3$, $S^3$ is parallelizable using quaternions. That is to say, given three linearly-independent vector fields forming a basis of the tangent space at some point of $S^3$, using quaternions it is possible to find three linearly-independent vector fields forming a basis of the tangent space at every other point of $S^3$. In other words, it is possible to set all of the points of a quaternionic $S^3$ in a smooth flowing motion at the same time, in any direction, without a fixed point, or a pole, or a singularity, or a cowlick. On the other hand, the tangent bundle of ${S^3}$ happens to be trivial: ${{\rm T}S^3 = S^3 \times{\mathrm{I\!R}}^3}$. This renders the tangent space at each point of ${S^3}$ to be isomorphic to ${{\mathrm{I\!R}}^3}$. Consequently, local experiences of the experimenters within ${S^3}$ are no different from those of their counterparts within ${{\mathrm{I\!R}}^3}$. The global topology of ${S^3}$, however, is clearly different from that of ${{\mathrm{I\!R}}^3}$ \cite{Disproof,IJTP}. In particular, the triviality of the bundle ${{\rm T}S^3}$ again means that ${S^3}$ is parallelizable. As a result, a global {\it anholonomic} frame can be defined on ${S^3}$ that fixes each of its points uniquely. Such a frame renders ${S^3}$ diffeomorphic to the set of all unit quaternions, as in (\ref{nonsin}). The properties of $S^3$ are thus uniquely captured by the properties of quaternions. 

Now, given two unit vectors ${\bf x}$ and ${\bf y}$ in ${\mathrm{I\!R}^3}$ and a rotation axis ${\bf r}$, each element of the set ${S^3}$ can be factorized into a product of corresponding bivectors ${{\bf J}({\bf x})}$ and ${{\bf J}({\bf y})}$ as follows:
\begin{align}
{\bf q}(\eta_{{\bf x}{\bf y}},\,{\bf r})&=-\,{\bf J}({\bf x})\,{\bf J}({\bf y}) \label{dep}\\ 
&=-\,(I\,{\bf x})\,(I\,{\bf y}) \\
&=-I^2\,{\bf x}\,{\bf y} \\
&={\bf x}\,{\bf y} \\
&=\,{\bf x}\cdot{\bf y}\,+\,{\bf x}\wedge{\bf y} \\
&=\cos(\,\eta_{{\bf x}{\bf y}})\,+\,{\bf J}({\bf r})\,\sin(\,\eta_{{\bf x}{\bf y}})\,, \label{new2}
\end{align}
where $I:={\bf e}_1\,{\bf e}_2\,{\bf e}_3$, with $I^2=-1$, is the standard trivector, ${\eta_{{\bf x}{\bf y}}}$ is the angle between ${\bf x}$ and ${\bf y}$, ${{\bf x}\,{\bf y}}$ is the geometric product between ${\bf x}$ and ${\bf y}$, ${{\bf x}\wedge{\bf y}}$ is the wedge product between ${\bf x}$ and ${\bf y}$, and ${{\bf J}({\bf r})}$ is identified with ${\frac{{\bf x}\wedge{\bf y}}{||{\bf x}\wedge{\bf y}||}}$. Comparing (\ref{nonsin}) and (\ref{new2}), we recognize that the rotation angle ${\psi}$ of the quaternion is twice the angle between the vectors ${\bf x}$ and ${\bf y}$:
\begin{equation}
\psi\,=\,2\,\eta_{{\bf x}{\bf y}}.
\end{equation}
As a result, the characteristic spinorial sign changes exhibited by the quaternions constituting the $S^3$ can be expressed as
\begin{equation}
{\bf q}(\eta_{{\bf x}{\bf y}}+\kappa\pi,\,{\bf r})\,=\,(-1)^{\kappa}\,{\bf q}(\eta_{{\bf x}{\bf y}},\,{\bf r})\,\;\;\text{for}\;\,\kappa=0,1,2,3,\dots \label{spinorial}
\end{equation}
This equation expresses a key relation that reduces the singlet correlations we observe in Nature to Dr.~Bertlmann's socks type classical correlations \cite{Bell-1987}, because, as we shall soon see, it forces the product ${{\mathscr A}{\mathscr B}}$ of the measurement results ${\mathscr A}=\pm$ and ${\mathscr B}=\pm$ observed, respectively, by Alice and Bob to fluctuate between the values ${{\mathscr A}{\mathscr B}=-1}$ and ${{\mathscr A}{\mathscr B}=+1}$ and vice versa. It thereby allows all four combinations of the results, ${{\mathscr A}{\mathscr B}=+\,+}$, ${+\,-}$, ${-\,+}$, and ${-\,-}$, necessary to produce the observed strong correlations between them. 

Incidentally, in the algebra $\mathrm{Cl}_{(3,0)}$ of a three-dimensional space the four-dimensional object $I\wedge{\bf v}$ is necessarily zero by definition. The vectors ${\bf v}$ in $\mathrm{Cl}_{(3,0)}$ are thus defined as the solutions of the equation $I\wedge{\bf v}=0$. Therefore, we can write
\begin{equation}
I{\bf v} = I\cdot{\bf v} + I \wedge {\bf v} = I\cdot{\bf v}.
\end{equation}
This choice is explained in Question 8 of Appendix~B in \cite{IEEE-1}. 

Next, let us recall that the measurement results $\mathscr{A}$ and $\mathscr{B}$ observed by Alice and Bob are events in spacetime. Within the spacetime defined by the line element (\ref{frw}), they are thus events in ${\mathrm{I\!R}\times\Sigma}$. Now, traditionally, in Bell-test experiments $\Sigma$ is implicitly identified with $\mathrm{I\!R}^3$. In other words, the three-dimensional physical space is implicitly modeled as $\mathrm{I\!R}^3$. Thus, traditionally, the results $\mathscr{A}$ and $\mathscr{B}$ observed by Alice and Bob are assumed to be events in ${\mathrm{I\!R}\times{\mathrm{I\!R}^3}}$. But in my model ${\mathrm{I\!R}^3}$ in ${\mathrm{I\!R}\times\mathrm{I\!R}^3}$ is replaced with $S^3$, and thus $S^3$ is taken to be a spacelike hypersurface in spacetime, and hence a surface of simultaneity. In other words, the results $\mathscr{A}$ and $\mathscr{B}$ are viewed as events in ${\mathrm{I\!R}\times S^3}$. But in the EPR-Bohm type experiments the observed results are necessarily equal-time events, otherwise called ``coincidence counts.'' Thus, in my model they are points in $S^3$ at the time of simultaneous measurements by Alice and Bob. These considerations, thanks to the decomposition (\ref{dep}), lead us to the following theorem.

\underbar{Theorem 1}: The quantum mechanical correlation predicted by the entangled singlet state can be understood as a classical, local, realistic, and deterministic correlation among the pairs of limiting scalar points of a quaternionic 3-sphere defined in (\ref{nonsin}), with the limiting scalar points defined by the functions
\begin{align}
S^3\ni\,{\mathscr A}({\bf a},\,{\lambda^k})\,&:=\lim_{{\bf s}_1\,\rightarrow\,{\bf a}}\left\{\,+\,{\bf q}(\eta_{{\bf a}{\bf s}_1},\,{\bf r}_1)\right\} \notag \\
&\equiv\lim_{{\bf s}_1\,\rightarrow\,{\bf a}}\left\{-\,{\bf D}({\bf a})\,{\bf L}({\bf s}_1,\,\lambda^k)\right\} \notag \\
&\xrightarrow[{\bf s}_1\,\to\,{\bf a}]\,\begin{cases}
+\,1\;\;\;\;\text{if} &\lambda^k\,=\,+\,1 \\
-\,1\;\;\;\;\text{if} &\lambda^k\,=\,-\,1
\end{cases} \Bigg\}\label{53}
\end{align}
and
\begin{align}
S^3\ni\,{\mathscr B}({\bf b},\,{\lambda^k})\,&:=\lim_{{\bf s}_2\,\rightarrow\,{\bf b}}\left\{\,-\,{\bf q}(\eta_{{\bf s}_2{\bf b}},\,{\bf r}_2)\right\} \notag \\
&\equiv\lim_{{\bf s}_2\,\rightarrow\,{\bf b}}\left\{+\,{\bf L}({\bf s}_2,\,\lambda^k)\,{\bf D}({\bf b})\right\} \notag \\
&\xrightarrow[{\bf s}_2\,\to\,{\bf b}]\,\begin{cases}
-\,1\;\;\;\;\text{if} &\lambda^k\,=\,+\,1 \\
+\,1\;\;\;\;\text{if} &\lambda^k\,=\,-\,1
\end{cases} \Bigg\}, \label{54}
\end{align}
where the bivectors ${-\,{\bf L}({\bf s}_1,\,\lambda^k)}$ and ${+\,{\bf L}({\bf s}_2,\,\lambda^k)}$ represent the two fermionic spins emerging from a common source that are subsequently detected (possibly at a space-like distance from each other) by two detector bivectors ${{\bf D}({\bf a})=I\cdot{\bf a}}$ and ${{\bf D}({\bf b})=I\cdot{\bf b}}$, freely chosen by Alice and Bob. I have also assumed the handedness ${\lambda^k}$ of ${S^3}$ to be a fair coin with 50/50 chance of being ${+1}$ or ${-\,1}$ at the moment of pair-creation, making the spinning bivector ${{\bf L}({\bf n},\,\lambda^k)}$ a random variable {\it relative} to any given detector bivector such as ${{\bf D}({\bf n})=I\cdot{\bf n}}$,
\begin{equation}
{\bf L}({\bf n},\,\lambda^k)\,=\,\lambda^k\,{\bf D}({\bf n})\,\,\Longleftrightarrow\,\,{\bf D}({\bf n})\,=\,\lambda^k\,{\bf L}({\bf n},\,\lambda^k)\,. \label{55}
\end{equation}

The next question is: What will be the value of the product ${{\mathscr A}{\mathscr B}}$ of these results {\it within} $S^3$? In other words, what will be the value of the product ${{\mathscr A}{\mathscr B}}$ when the results ${\mathscr A}$ and ${\mathscr B}$ are observed by Alice and Bob separately but simultaneously in ``coincidence counts'' \cite{RSOS}? We can work out the value of the product ${{\mathscr A}{\mathscr B}}$ within $S^3$ from the definitions (\ref{53}) and (\ref{54}) and the ``product of limits equal to limits of product'' rule:
\begin{align}
S^3&\ni\,{\mathscr A}{\mathscr B}({\bf a},\,{\bf b},\,{\lambda^k})\,=\,{\mathscr A}({\bf a},\,{\lambda^k})\,{\mathscr B}({\bf b},\,{\lambda^k}) \label{16a} \\
&=\,\left[\lim_{{\bf s}_1\,\rightarrow\,{\bf a}}\left\{\,+\,{\bf q}(\eta_{{\bf a}{\bf s}_1},\,{\bf r}_1)\right\}\right]\left[\lim_{{\bf s}_2\,\rightarrow\,{\bf b}}\left\{\,-\,{\bf q}(\eta_{{\bf s}_2{\bf b}},\,{\bf r}_2)\right\}\right] \\
&=\,\lim_{\substack{{\bf s}_1\,\rightarrow\,{\bf a} \\ {\bf s}_2\,\rightarrow\,{\bf b}}}\left\{\,-\,{\bf q}(\eta_{{\bf a}{\bf s}_1},\,{\bf r}_1)\,{\bf q}(\eta_{{\bf s}_2{\bf b}},\,{\bf r}_2)\right\} \\
&=\,\lim_{\substack{{\bf s}_1\,\rightarrow\,{\bf a} \\ {\bf s}_2\,\rightarrow\,{\bf b}}}\left\{\,-\,{\bf q}(\eta_{{\bf u}{\bf v}},\,{\bf r}_{0})\right\} \label{35}\\
&=\,-1, \label{36}
\end{align}
where, for the moment, I have assumed that ${\bf s}_1\not={\bf s}_2$, giving
\begin{align}
\eta_{{\bf u}{\bf v}}:=\cos^{-1}&\big\{({\bf a}\cdot{\bf s}_1)({\bf s}_2\cdot{\bf b}) \notag \\
&\;\;\;-({\bf a}\cdot{\bf s}_2)({\bf s}_1\cdot{\bf b})+({\bf a}\cdot{\bf b})({\bf s}_1\cdot{\bf s}_2)\big\} \label{38}
\end{align}
and
\begin{equation}
{\bf r}_{0}=\frac{\scriptstyle{({\bf a}\cdot{\bf s}_1)({\bf s}_2\times{\bf b})\,+\,({\bf s}_2\cdot{\bf b})({\bf a}\times{\bf s}_1)\,-\,({\bf a}\times{\bf s}_1)\times({\bf s}_2\times{\bf b})}}{\scriptstyle{||\,({\bf a}\cdot{\bf s}_1)({\bf s}_2\times{\bf b})\,+\,({\bf s}_2\cdot{\bf b})({\bf a}\times{\bf s}_1)\,-\,({\bf a}\times{\bf s}_1)\times({\bf s}_2\times{\bf b})\,||}}. \label{null}
\end{equation}
That the product of the two remote quaternions ${{\bf q}(\eta_{{\bf a}{\bf s}_1},\,{\bf r}_1)}$ and ${{\bf q}(\eta_{{\bf s}_2{\bf b}},\,{\bf r}_2)}$ is yet another quaternion ${{\bf q}(\eta_{{\bf u}{\bf v}},\,{\bf r}_{0})}$ is not surprising, because the set ${S^3}$ defined in (\ref{nonsin}) is known to remain closed under multiplication. A product of any number of quaternions will result in yet another quaternion belonging to ${S^3}$. More importantly for our hypothesis, the product ${{\mathscr A}{\mathscr B}({\bf a},\,{\bf b},\,{\lambda^k})}$ is again a limiting scalar point, ${-1}$ in this case, of the quaternion ${-\,{\bf q}(\eta_{{\bf u}{\bf v}},\,{\bf r}_{0})}$ that also belongs to ${S^3}$.

The result (\ref{36}), namely ${{\mathscr A}{\mathscr B}=-1}$, suggests that if Alice finds spin to be ``up'' at her station, then Bob is guaranteed to find spin to be ``down'' at his station, precisely mimicking the perfect anti-correlation observed in Dr.~Bertlmann's socks type correlations. This may give the wrong impression that the product ${{\mathscr A}{\mathscr B}}$ of the results observed by Alice and Bob will always remain at the fixed value of ${-1}$. But because of the spinorial sign changes described in (\ref{spinorial}), which any quaternion in ${S^3}$ --- including the quaternion ${-\,{\bf q}(\eta_{{\bf u}{\bf v}},\,{\bf r}_{0})}$ appearing in Eq.~(\ref{35}) as well as those appearing in the definitions (\ref{53}) and (\ref{54}) of the individual measurement results  ${\mathscr A}$ and ${\mathscr B}$ --- must respect, the value of the product ${{\mathscr A}{\mathscr B}}$ will be altered. Moreover, variations in the detector directions ${\bf a}$ and ${\bf b}$ will induce variations in the angle ${\eta_{{\bf u}{\bf v}}}$ defined in (\ref{38}), which can be expressed as ${\eta_{{\bf u}{\bf v}}\rightarrow\eta_{{\bf u}{\bf v}}+\delta}$. For variation ${\delta=\kappa\pi}$, the quaternion ${-\,{\bf q}(\eta_{{\bf u}{\bf v}},\,{\bf r}_{0})}$ appearing in Eq.~(\ref{35}) will then change its sign from ${-\,{\bf q}(\eta_{{\bf u}{\bf v}},\,{\bf r}_{0})}$ to ${+\,{\bf q}(\eta_{{\bf u}{\bf v}},\,{\bf r}_{0})}$ for odd ${\kappa}$. As a result, the value of the product ${{\mathscr A}{\mathscr B}}$ will change from ${-1}$ to ${+1}$ for odd ${\kappa}$. We would thus have our cake ({\it i.e.}, Dr.~Bertlmann's socks type local-realistic interpretation of the correlations) and eat it too --- {\it i.e.}, have the value of the product ${{\mathscr A}{\mathscr B}}$ fluctuate between ${-1}$ and ${+1}$:
\begin{equation}
{\mathscr A}{\mathscr B}\in\{-1,\,+1\}.
\end{equation}
In other words, all four possible combinations of outcomes, ${+\,+}$, ${+\,-}$, ${-\,+}$, and ${-\,-}$, will be observed by Alice and Bob despite the correlations being Dr.~Bertlmann's socks type.

But that is only a necessary part of the singlet correlations. In deriving the value $-1$ of the product ${{\mathscr A}{\mathscr B}}$ in (\ref{36}) we have assumed ${\bf s}_1\not={\bf s}_2$. That assumption, however, violates the conservation of zero spin angular momentum of the singlet,
\begin{align}
-\,{\bf L}({\bf s}_1,\,\lambda^k)+{\bf L}({\bf s}_2,\,\lambda^k)=0&\Longleftrightarrow {\bf L}({\bf s}_1,\,\lambda^k)={\bf L}({\bf s}_2,\,\lambda^k) \notag \\
&\Longleftrightarrow {\bf s}_1=\,{\bf s}_2\,\equiv\,{\bf s}\,, \label{56a}
\end{align}
which necessarily holds during the free evolution of the spins ${-\,{\bf L}({\bf s}_1,\,\lambda^k)}$ and ${+\,{\bf L}({\bf s}_2,\,\lambda^k)}$ from the source $\pi^0$ until their detections by ${{\bf D}({\bf a})}$ and ${{\bf D}({\bf b})}$, as specified in (\ref{53}) and (\ref{54}). This condition, in the light of the analog (\ref{50a}) of the Pauli identity discussed below, is also equivalent to the condition
\begin{equation}
{\bf L}({\bf s}_1,\,\lambda^k)\,{\bf L}({\bf s}_2,\,\lambda^k)=\left\{\,{\bf L}({\bf s},\,\lambda^k)\right\}^2=\,{\bf L}^2({\bf s},\,\lambda^k)=-1\,. \label{566a}
\end{equation}
As proved in Section VIII of \cite{IEEE-1} and Appendix~A of \cite{IEEE-2}, in the context of EPR-Bohm experiments this algebraic condition for the conservation spin angular momentum is equivalent to the M\"obius-like twists in the fiber geometry of ${S^3}$. Thus, the condition 
${{\bf s}_1={\bf s}_2}$ is a part of the very geometry of the physical space $S^3$ within which we are inescapably confined to perform our experiments. Therefore, the result (\ref{36}), which can be valid within the physical space modeled as ${\mathrm{I\!R}^3}$, cannot possibly be valid within the physical space modeled as $S^3$.

It is important to note that, even though the conservation of zero spin angular momentum of the singlet implies ${\bf s}_1={\bf s}_2$, the {\it physical} sense of ${\bf s}_1$ proceeding towards Alice's detection process (or measurement interaction) defined in (\ref{53}) and ${\bf s}_2$ proceeding towards Bob's detection process (or measurement interaction) defined in (\ref{54}) remain unchanged. Moreover, for ${{\bf s}_1={\bf s}_2}$ the angle ${\eta_{{\bf u}{\bf v}}}$ defined in (\ref{38}) reduces to the angle ${\eta_{{\bf a}{\bf b}}}$ between the detector directions ${\bf a}$ and ${\bf b}$. As a result, for ${{\bf s}_1={\bf s}_2}$ the product of the measurement results reduces to
\begin{align}
{\mathscr A}({\bf a},\,{\lambda^k})\,&{\mathscr B}({\bf b},\,{\lambda^k})\,\longrightarrow\lim_{\substack{{\bf s}_1\,\rightarrow\,{\bf a} \\ {\bf s}_2\,\rightarrow\,{\bf b}}}\left\{\,-\,{\bf q}(\eta_{{\bf a}{\bf b}},\,{\bf r}_{0})\right\} \notag \\
&=\lim_{\substack{{\bf s}_1\,\rightarrow\,{\bf a} \\ {\bf s}_2\,\rightarrow\,{\bf b}}}\left\{-\cos(\,\eta_{{\bf a}{\bf b}})-{\bf L}({\bf r}_{0},\,\lambda^k)\,\sin(\,\eta_{{\bf a}{\bf b}})\right\}. \label{notpro}
\end{align}
Note that, so far, we have not taken the limits ${\bf s}_1\to {\bf a}$ and ${\bf s}_2\to {\bf b}$. But for ${{\bf s}_1={\bf s}_2}$ the only quantity in (\ref{notpro}) that still depends on the spin directions ${\bf s}_1$ and ${\bf s}_2$ is the rotation axis vector ${\bf r}_0$, which, for ${{\bf s}_1={\bf s}_2}$, using (\ref{null}), works out to be
\begin{equation}
{\bf r}_{0}=\frac{\scriptstyle{({\bf a}\cdot{\bf s}_1)({\bf s}_2\times{\bf b})\,+\,({\bf s}_2\cdot{\bf b})({\bf a}\times{\bf s}_1)\,-\,({\bf a}\times{\bf s}_1)\times({\bf s}_2\times{\bf b})}}{\scriptstyle{\sin\left(\,\eta_{{\bf a}{\bf b}}\right)}}. \label{null-3}
\end{equation}
Consequently, in the simultaneous limits ${\bf s}_1\to {\bf a}$ and ${\bf s}_2\to {\bf b}$ characterising the measurement interactions defined in (\ref{53}) and (\ref{54}) (during which spin angular momentum is no longer conserved and therefore the condition ${{\bf s}_1={\bf s}_2}$ does not hold), the rotation axis vector ${\bf r}_0$ in (\ref{notpro}) reduces to a null vector:
\begin{equation}
\lim_{\substack{{\bf s}_1\,\rightarrow\,{\bf a} \\ {\bf s}_2\,\rightarrow\,{\bf b}}}\left\{\,{\bf r}_{0}\,\right\}=\vec{\,0}. \label{null-2}
\end{equation}
As a result, the bivector in (\ref{notpro}) also reduces to a null bivector:
\begin{align}
\lim_{\substack{{\bf s}_1\,\rightarrow\,{\bf a} \\ {\bf s}_2\,\rightarrow\,{\bf b}}}\left\{{\bf L}({\bf r}_{0},\,\lambda^k)\,\sin(\,\eta_{{\bf a}{\bf b}})\right\}&={\bf L}(\vec{\bf\,0},\,\lambda^k)\,\sin(\,\eta_{{\bf a}{\bf b}}) \notag \\
&=(J\cdot\vec{\bf\,0}\,)\,\sin(\,\eta_{{\bf a}{\bf b}}),
\end{align}
where $J$ is a trivector representing the volume form on $S^3$. Consequently, the product ${{\mathscr A}{\mathscr B}}$ in (\ref{notpro}) tends to $-\cos(\,\eta_{{\bf a}{\bf b}})$:
\begin{align}
{\mathscr A}({\bf a},\,{\lambda^k})&\,{\mathscr B}({\bf b},\,{\lambda^k})\,\longrightarrow\lim_{\substack{{\bf s}_1\,\rightarrow\,{\bf a} \\ {\bf s}_2\,\rightarrow\,{\bf b}}}\left\{\,-\,{\bf q}(\eta_{{\bf a}{\bf b}},\,{\bf r}_{0})\right\} \notag \\
&=\lim_{\substack{{\bf s}_1\,\rightarrow\,{\bf a} \\ {\bf s}_2\,\rightarrow\,{\bf b}}}\left\{-\cos(\,\eta_{{\bf a}{\bf b}})-{\bf L}({\bf r}_{0},\,\lambda^k)\,\sin(\,\eta_{{\bf a}{\bf b}})\right\}, \notag \\
&=\,-\cos(\,\eta_{{\bf a}{\bf b}})-{\bf L}(\vec{\bf \,0},\,\lambda^k)\,\sin(\,\eta_{{\bf a}{\bf b}}),
\notag \\
&=\,-\cos(\,\eta_{{\bf a}{\bf b}})\,-\,0\,.
\end{align}
Evidently, this tendency of ${{\mathscr A}{\mathscr B}}$ holds for each run of the experiment. Consequently, using 
the universally accepted definition of the correlations function used in the Bell-test experiments, the correlation between the results ${{\mathscr A}({\bf a},\,{\lambda^k})}$ and ${{\mathscr B}({\bf b},\,{\lambda^k})}$ observed by Alice and Bob work out to give
\begin{subequations}
\begin{align}
&{\cal E}_{\rm L.R.}({\bf a},\,{\bf b})=\!\lim_{\,n\,\gg\,1}\left[\frac{1}{n}\sum_{k\,=\,1}^{n}\,{\mathscr A}({\bf a},\,{\lambda}^k)\;{\mathscr B}({\bf b},\,{\lambda}^k)\right] \label{57a} \\
&\;\;\;\;=\lim_{\,n\,\gg\,1}\left[\frac{1}{n}\sum_{k\,=\,1}^{n}\,\lim_{\substack{{\bf s}_1\,\rightarrow\,{\bf a} \\ {\bf s}_2\,\rightarrow\,{\bf b}}}\left\{\,-\,{\bf q}(\eta_{{\bf a}{\bf b}},\,{\bf r}_{0})\right\}\right] \label{60a}\\
&\;\;\;\;=\lim_{\,n\,\gg\,1}\left[\frac{1}{n}\sum_{k\,=\,1}^{n}\{-\cos(\,\eta_{{\bf a}{\bf b}})-{\bf L}(\vec{\bf \,0},\,\lambda^k)\,\sin(\,\eta_{{\bf a}{\bf b}})\}\right] \label{61a} \\
&\;\;\;\;=\,-\cos(\,\eta_{{\bf a}{\bf b}})\,-\!\lim_{\,n\,\gg\,1}\left[\frac{1}{n}\sum_{k\,=\,1}^{n}\,{\bf L}(\vec{\bf \,0},\,\lambda^k)\,\sin(\,\eta_{{\bf a}{\bf b}})\,\right]\label{63a}\\
&\;\;\;\;=\,-\cos(\,\eta_{{\bf a}{\bf b}})\,-\,0\,. \label{65a}
\end{align}
\end{subequations}
This corroborates my hypothesis that the observed singlet correlations are correlations among the limiting scalar points ${{\mathscr A}({\bf a},\,{\lambda})=\pm1}$ and ${{\mathscr B}({\bf b},\,{\lambda})\pm1}$ of a quaternionic 3-sphere. It is also worth noting that the above derivation of (\ref{65a}) is just one of several different ways Theorem~1 is proved in \cite{Disproof,IJTP,RSOS,IEEE-1,IEEE-2}.

Despite this inevitable result, replacing the global topology of the physical space from $\mathrm{I\!R}^3$ to $S^3$ may seem pointless, because locally, in the topological sense, $S^3$ is isomorphic to $\mathrm{I\!R}^3$, similarly to how Earth ($S^2$) is isomorphic to  ${\mathrm{I\!R}^2}$. At each point of $S^3$ the tangent space is simply $\mathrm{I\!R}^3$. Consequently, in local experiments the normalized directions such as ${\bf a}$ and ${\bf b}$ can certainly be taken to be from $S^2\in\mathrm{I\!R}^3$. But that does not rule out the possibility that they are, in fact, embedded in a quaternionic 3-sphere, as I have proposed in \cite{IEEE-2}. An {\it a~priori} denial of the possibility of a global $S^3$ nature of space would be analogous to a denial of the spherical nature of Earth.

\section{Point-by-point response to the critique} \label{II}

Unfortunately, the critique in \cite{Gill-Ieee} ignores the above 3-sphere model and its physical significance entirely. Instead of appreciating that the model is based on the orientation $\lambda=\pm1$ of a closed and compact physical space $S^3$, it insists on reinterpreting it as a hidden variable model based on a detached binary number $\pm1$ within a flat and non-compact space ${\mathrm{I\!R}}^3$.

While there are too many incorrect claims throughout the critique, in this Section I focus on those that are significant.

\subsection{Concerning one of Bell's assumptions}\label{B}

Ironically, the critique itself makes one of the implicit and unjustifiable assumptions of Bell's theorem quite explicit in its statement of the theorem \cite{Gill-Ieee}:
\begin{quote}
In the {\it language of probability theory}, the mathematical core of Bell's original proof of his theorem is the assertion that one cannot find a single probability space on which are defined random variables $X_{\mathbf a}$ and $Y_{\mathbf b}$ taking values in the set $\{-1, +1\}$, for all ${\mathbf a}$, ${\mathbf b}$, unit vectors in $\mathbb R^3$,  and such that
$$
{\mathbb E}(X_{\mathbf a} Y_{\mathbf b}) ~=~ -{\mathbf a} \cdot {\mathbf b}
\eqno(1)
$$
for all $\mathbf a$, $\mathbf b$. 
Moreover, the expectation values of $X_{\mathbf a}$ and $Y_{\mathbf b}$ are all zero.
\end{quote}
But why must we assume unit vectors ${\bf a}$, ${\bf b}$ to be in ${\mathrm{I\!R}^3}$? As I have summarized in Section \ref{I}, there are both theoretical and observational reasons that compel us to model physical space as a closed and compact quaternionic 3-sphere, or $S^3$, instead of a flat Euclidean space ${\mathrm{I\!R}^3}$, both being admissible spatial parts of one of the most well known cosmological solutions of Einstein's field equations of general relativity. Moreover, as explained in several of my papers since 2007 \cite{Disproof,IJTP,RSOS,IEEE-1,IEEE-2}, the correct language to model $S^3$ as physical space is Geometric Algebra, not vector ``algebra.'' This implies, in particular, that the directions ${\bf a}$ and ${\bf b}$ freely chosen by Alice and Bob to perform their experiments must be solutions of the equation
\begin{equation}
I\wedge{\bf v} = 0,
\end{equation}
where $I$ is a volume form on the physical space $S^3$ and ${\bf v}$ is any vector built from the orthogonal directions $\{{\bf e}_x,\,{\bf e}_y,\,{\bf e}_z\}$ that characterize the Clifford algebra ${\mathrm{Cl}_{(3,0)}}$. Once the physical space is modeled as $S^3$ instead of $\mathrm{I\!R}^3$ in the manner explained in Section \ref{I} and characterized using the powerful language of Geometric Algebra, the correlations between the measurement results ${\mathscr{A}_{\bf a}}$ and ${\mathscr{B}_{\bf b}}$ observed by Alice and Bob inevitably turn out to be ${\cal E}({\mathscr{A}_{\bf a}}{\mathscr{B}_{\bf b}})=-{\bf a}\cdot{\bf b}$ as proven in \cite{IEEE-2}.

On the other hand, the standard interpretation of Bell's theorem adhered to in \cite{Gill-Ieee} is recovered in the flat geometry of ${{\mathrm{I\!R}^3}}$. The ${S^3}$ model becomes conducive to the traditional interpretation of the theorem when the algebraic, geometrical and topological properties of the compactified physical space ${S^3}$ are ignored. In that case the upper bound of 2 on the Bell-CHSH inequalities is respected. Thus, the results presented in \cite{Disproof,IJTP,RSOS,IEEE-1,IEEE-2} do not conflict with the standard interpretation of Bell's theorem outright but rather reproduces that interpretation as a special case in the flat geometry ${{\mathrm{I\!R}}^3}$ of the physical space. This is discussed in more detail in Section X of \cite{IEEE-1}.

\subsection{Concerning a denial of Bell's assumption}

In the last sentence of Introduction the critique \cite{Gill-Ieee} claims that
\begin{quote}
Bell does not take account of the geometry of space because his argument, on the side of local realism, does not depend on it in any way whatsoever.
\end{quote}
But this claim is immediately contradicted in the critique in the very paragraph that follows the sentence quoted above. As I just discussed in the previous subsection, the critique explicitly assumes the geometry of physical space to be ${\mathrm{I\!R}^3}$ in its statement of Bell's theorem. Indeed, it is not possible for us to escape the geometry of physical space while performing our experiments. For this reason, nowhere in his writings has Bell stated that his theorem holds independently of the geometry of physical space. In fact, Bell's proposed local-realistic framework {\it does} assume implicitly that physical space in which we are confined to perform our experiments is modeled as $\mathrm{I\!R}^3$, as the critique does in the first paragraph of its Section~II. It is unfortunate that this assumption is usually not made explicit in the literature on the subject. A deeper reflection on how the physical space is modeled in analyzing the Bell-test experiments is necessary to uncover this implicit assumption. In the analyses of such experiments ordinary vector algebra (which does not, in fact, form an algebra at all) within $\mathrm{I\!R}^3$ is implicitly assumed. On the other hand, I have modeled the physical space as a quaternionic 3-sphere using Geometric Algebra \cite{Clifford}. The fact that I have been able to reproduce the strong singlet correlations by modeling the physical space as a quaternionic 3-sphere is a confirmatory evidence that the strategy to relax the unwarranted implicit assumption built-in Bell's theorem has been successful.

\subsection{Concerning Local causality of the model}

In its Introduction, the critique \cite{Gill-Ieee} claims that ``Christian's idea that quantum correlations are explained by the geometry of space might seem appealing, but ... such an explanation would not be ``local'' in any meaningful sense.'' However, we are not at liberty to guess what is or is not a meaningful sense of locality. Einstein and Bell have given us a very precise notion of local causality and I have strictly adhered to that notion throughout my work on the subject. As is well known, a violation of relativistic local causality can be separated into two distinct parts: (1) a signalling non-locality incompatible with general relativity, and (2) a non-signalling non-locality compatible with general relativity. These two distinct parts are captured by Bell in his definitions ${{\mathscr A}({\bf a},\,\lambda)}$ and ${{\mathscr B}({\bf b},\,\lambda)}$ of local measurement functions for any given initial state ${\lambda}$ of a given physical system. This separates relativistic local causality into independence of the parameter ${\bf a}$ from ${\bf b}$ (and vice versa) preserving signalling locality, and independence of the outcome ${\mathscr A}$ from ${\mathscr B}$ (and vice versa) preserving non-signalling locality, in any EPR-Bohm type experiment.

In the model presented in \cite{Disproof,IJTP,RSOS,IEEE-1,IEEE-2} and the previous section the question of signalling non-locality does not arise because the quaternionic 3-sphere on which it is based is a part of the solution of Einstein's field equations of general relativity. And the question of non-signalling non-locality is also implicitly addressed within the model by recognizing that the measurement functions (\ref{53}) and (\ref{54}) define manifestly local-realistic functions. Apart from the hidden variable ${\lambda}$, the result ${{\mathscr A}=\pm1}$ depends {\it only} on the measurement direction ${\bf a}$, chosen freely by Alice, regardless of Bob's actions. And similarly, apart from the hidden variable ${\lambda}$, the result ${{\mathscr B}=\pm1}$ depends {\it only} on the measurement direction ${\bf b}$, chosen freely by Bob, regardless of Alice's actions. In particular, the function ${{\mathscr A}({\bf a},\,\lambda)}$ {\it does not} depend on ${\bf b}$ or ${\mathscr B}$ and the function ${{\mathscr B}({\bf b},\,\lambda)}$ {\it does not} depend on ${\bf a}$ or ${\mathscr A}$. Moreover, the hidden variable ${\lambda}$ does not depend on either ${\bf a}$, ${\bf b}$, ${\mathscr A}$, or ${\mathscr B}$.

Unimpressed by this, the critique \cite{Gill-Ieee} continues its claim:
\begin{quote}
Christian seems to see the local spatial coordinate system of Alice being the mirror image of Bob's, the two orientations being determined completely at random, again and again! However, in modern accounts of Bell's theorem, angles and orientations play no role whatsoever. The new generation of loophole-free Bell experiments ... measure correlations between four binary variables: two binary inputs and two binary outputs; one input and output at each of two distant locations.
\end{quote}
These comments provide a clear evidence that the critique in \cite{Gill-Ieee} is based on a mistaken understanding of what is meant by the orientation $\lambda$ of the 3-sphere. The orientation $\lambda$ of $S^3$ does not concern the local spatial coordinate systems of Alice and Bob. It describes the handedness of the physical space $S^3$ itself. It specifies whether the 3-sphere is inside-out or outside-out\footnote{A good analogy of this is an ordinary hand-glove. If the outside of a hand-glove is right-handed, then pulling it inside out will make it left-handed.} with respect to the detectors in a given run of the experiment, with 50/50 chance. The value of $\lambda$ is fixed for the detectors ${{\bf D}({\bf a})=I\cdot{\bf a}}$ and ${{\bf D}({\bf b})=I\cdot{\bf b}}$, chosen freely by Alice and Bob, for all runs of the experiment. But it is not fixed for the spins ${-\,{\bf L}({\bf s}_1,\,\lambda)}$ and ${+\,{\bf L}({\bf s}_2,\,\lambda)}$ emerging from the source. Thus, $\lambda$ plays a role of a hidden variable, or an initial state of the system, {\it relative} to the fixed handedness of the detectors. Since it plays the role of a hidden variable, Alice and Bob need not be concerned about it at all. All they need to worry about are four binary variables: two binary inputs ``${\bf a}$'' and ``${\bf b}$'', and two binary outputs ``${\mathscr A}$'' and ``${\mathscr B}$''; one input and one output at each of the two distant locations.

\subsection{Concerning Equations (32) and (33) of \cite{IEEE-2}}

In Section III of the critique, in the paragraph before the one containing equation (4), it is stated, incorrectly, that in \cite{IEEE-2} the bivectors ${\bf L}({\bf a},\lambda)$, ${\bf L}({\bf b},\lambda)$, ${\bf D}({\bf a})$, and ${\bf D}({\bf b})$ are introduced by equations (32) and (33). This incorrect starting point is the reason behind much of the confusion manifest in the critique. 

In fact, the spin bivectors ${\bf L}({\bf s},\lambda)$ in \cite{IEEE-2} are introduced by the bivector subalgebra (28), and the detector bivectors ${\bf D}({\bf n})$ are introduced by the bivector subalgebra (31); namely, by
\begin{equation}
L_{i}(\lambda)\,L_{j}(\lambda) \,=\,-\,\delta_{ij}\,-\,\sum_{k}\,\epsilon_{ijk}\,L_{k}(\lambda) \label{wh-o8899}
\end{equation}
and
\begin{equation}
D_{i}\,D_{j}\,=\,-\,\delta_{ij}\,-\,\sum_{k}\,\epsilon_{ijk}\,D_{k}\,, \label{wh-o8}
\end{equation}
respectively. In other words, the spin bivectors ${\bf L}({\bf s},\lambda)$ and the detector bivectors ${\bf D}({\bf n})$ are introduced using two {\it different} basis vectors, because of the experimental prerequisite that the detectors are located at remote stations at spacelike distance from each other, whereas the spins originate at the central source,  independently of the detectors. Indeed, it is explicitly stated in \cite{IEEE-2} that the basis bivectors for the spins,
\begin{equation}
L_i(\lambda)=J\cdot{\bf e}_i'\,, \label{228a}
\end{equation}
are defined in terms of the basis vectors ${\{{\bf e}_1',\,{\bf e}_2',\,{\bf e}_3'\}}$ and the corresponding trivector $J={\bf e}_1'{\bf e}_2'{\bf e}_3'\,$; and the basis bivectors for the detectors,
\begin{equation}
D_i=I\cdot{\bf e}_i\,, \label{229a}
\end{equation}
are defined in terms of the basis vectors ${\{{\bf e}_1,\,{\bf e}_2,\,{\bf e}_3\}}$ and the corresponding trivector $I={\bf e}_1{\bf e}_2{\bf e}_3$. It is surprising that these explicit definitions of $L_i(\lambda)$ and $D_i$ are missed in \cite{Gill-Ieee}. It is not easy to notice relations (32) and (33) in \cite{IEEE-2} but miss the explicit definitions (28) and (31) written in the same column.

In any case, once the above definitions (\ref{228a}) and (\ref{229a}) of $L_i(\lambda)$ and $D_i$ specifying two different bivector basis are {\it not} missed, with $\lambda$ being the orientation of $S^3$, then, as elaborated in Question 13 of Appendix~B in \cite{IEEE-2} leading up to Eq.~(74), the basis bivectors $L_i(\lambda)$ and $D_i$ are clearly related by $\lambda$ as
\begin{equation}
L_{i}(\lambda)\,=\,\lambda\,D_{i}\;\;\Longleftrightarrow\;\;
D_{i}\,=\,\lambda\,L_{i}(\lambda)\,.\label{1237} 
\end{equation}
Contracting on both sides of (\ref{1237}) with the components $n^i$ of an arbitrary unit vector ${\bf n}$ then gives the relation (\ref{55}) stated above, and contracting the bivector subalgebra defined in (\ref{wh-o8899}) above on both sides with the components $a^i$ and $b^j$ of arbitrary unit vectors ${\bf a}$ and ${\bf b}$ gives the Pauli identity,
\begin{equation}
{\bf L}({\bf a},\,\lambda)\,{\bf L}({\bf b},\,\lambda)\,=\,-\,{\bf a}\cdot{\bf b}\,-\,{\bf L}({\bf a}\times{\bf b},\,\lambda), \label{50a}
\end{equation}
with unit ${\mathbf{L}(\mathbf{a},\,\lambda):=a^iL_i(\lambda)}$ and unit ${\mathbf{L}(\mathbf{b},\,\lambda):=b^jL_j(\lambda)}$.

The above definitions allow the spin bivectors ${\bf L}({\bf a},\lambda)$ and ${\bf L}({\bf b},\lambda)$ to relate to the detector bivectors ${\bf D}({\bf a})$ and ${\bf D}({\bf b})$ by the orientation $\lambda$ of $S^3$ at the time of their measurements, as specified by equations (32) and (33) of \cite{IEEE-2}. The critique, however, ignores these definitions and insists on interpreting the contingent relations (32) and (33) between the spin bivectors and detector bivectors as the definition of the spin bivectors. Moreover, the critique's interpretation is also physically incorrect. For when they emerge from the common source the spins would be spinning about the direction ${\bf s}$ that would have no prior relation to what directions ${\bf a}$ and ${\bf b}$ Alice and Bob may have chosen to perform their measurements. And even if we go along with the critique's insistence on focusing on the contingent relations (32) and (33) as a piece of mathematics, the internal consistency of the 3-sphere model is robust enough to prevent the critique's strategy from succeeding, as I now demonstrate. 

The critique begins by identifying ${{\bf L}({\bf a},\,\lambda)}$ with $\lambda\,I\cdot{\bf a}$ and ${{\bf L}({\bf b},\,\lambda)}$ with $\lambda\,I\cdot{\bf b}$, but neglects to make the identification
\begin{equation}
{\bf L}({\bf a}\times{\bf b},\,\lambda)=\lambda\,I\cdot({\bf a}\times{\bf b}),
\end{equation}
which is demanded by the geometrical consistency of vectors within $S^2\subset\mathrm{I\!R}^3$. The critique \cite{Gill-Ieee} then claims:
\begin{quote}
It follows directly from Christian’s (32) and (33) that
\begin{equation}
\mathbf L(\mathbf a, \lambda) \mathbf L(\mathbf b, \lambda) = \lambda^2 I^2 \mathbf a \mathbf b = - \mathbf a \mathbf b, \tag{4} \label{gillwrong}
\end{equation}
which does not depend on $\lambda$ at all.
\end{quote}
But that is not correct. The RHS of the critique's Eq.~(\ref{gillwrong}) is not independent of $\lambda$. In fact, $\lambda$ is implicit in the product ${-{\bf a}{\bf b}=-{\bf a}\cdot{\bf b}-{\bf a}\wedge{\bf b}}$, not absent from it.  It would be a meaningless equation if its LHS “depended” on $\lambda$ while its RHS did not. Moreover, to begin with, Eqs.~(32) and (33) of \cite{IEEE-2} do not specify what the geometric product of ${{\bf L}({\bf a},\,\lambda)}$ with ${{\bf L}({\bf b},\,\lambda)}$ is. Eq.~(29) of \cite{IEEE-2} does. Thus, Eq.~(\ref{gillwrong}) of \cite{Gill-Ieee} is rather presumptuous. And why must we stop at the first step in Eq.~(\ref{gillwrong})? Why not continue the derivation of the product by recalling from the definition (\ref{228a}) above that, by definition, ${{\bf L}({\bf a},\,\lambda)=J\cdot{\bf a}}$ and therefore ${J\cdot{\bf a}=\lambda\,I\cdot{\bf a}}$ if we follow the critique's absolute identification, or, equivalently,
\begin{equation}
J=\lambda\,I\Longleftrightarrow\,I=\lambda\,J,
\end{equation}
which then gives
\begin{align}
{\bf L}({\bf a},\,\lambda)\,{\bf L}({\bf b},\,\lambda)\,&=\,-\,{\bf a}{\bf b} \\
&=\,-\,{\bf a}\cdot{\bf b}\,-\,{\bf a}\wedge{\bf b} \\
&=\,-\,{\bf a}\cdot{\bf b}\,-\,J\cdot({\bf a}\times{\bf b}) \\
&=\,-\,{\bf a}\cdot{\bf b}\,-\,\lambda\,I\cdot({\bf a}\times{\bf b}) \\
&=\,-\,{\bf a}\cdot{\bf b}\,-\,{\bf L}({\bf a}\times{\bf b},\,\lambda)\,.
\end{align}
But this is precisely Eq.~(29) of \cite{IEEE-2} or (\ref{50a}) above, and its RHS does ``depend'' on $\lambda$, contrary to the critique's claim. In other words, Eqs.~(32) and (33) of \cite{IEEE-2} {\it do not} contradict its Eq.~(29), and consequently the critique's argument and strategy fail.

Unfortunately, the critique's conceptual mistake here is even more serious. It treats the orientation $\lambda$ of $S^3$ as if it were an argument of the spin bivector ${\bf L}({\bf s},\,\lambda)$ by itself without reference to the detector bivector ${\bf D}({\bf n})$. But $\lambda$ is not just a number or an argument of ${\bf L}({\bf s},\,\lambda)$ by itself. It represents the handedness of the spin bivector ${\bf L}({\bf n},\,\lambda)$ {\it relative} to that of the detector bivector ${\bf D}({\bf n})$, and vice versa, as specified in Eq.~(\ref{55}) above. Thus, between ${\bf L}({\bf s},\,\lambda)$ and ${\bf D}({\bf n})$, $\lambda$ has only {\it relative} significance, and $\lambda$ for the spin ${\bf L}({\bf s},\,\lambda)$ is meaningful only with respect to the detector ${\bf D}({\bf n})$, and vice versa. Consequently, using the identity (\ref{50a}) and the relations (\ref{55}), the products of two spin bivectors can be evaluated as
\begin{align}
{\bf L}({\bf a},\,{\lambda}=+1)\;{\bf L}({\bf b},\,&{\lambda}=+1) \notag \\
&=\,-\,{\bf a}\cdot{\bf b}\,-\,{\bf L}({\bf a}\times{\bf b},\,\lambda=+1) \notag \\
&=-\,{\bf a}\cdot{\bf b}\,-\,{\bf D}({\bf a}\times{\bf b}) \notag \\
&={\bf D}({\bf a})\;{\bf D}({\bf b})
\end{align}
and
\begin{align}
{\bf L}({\bf a},\,{\lambda}=-1)\;{\bf L}({\bf b},\,&{\lambda}=-1) \notag \\
&=\,-\,{\bf a}\cdot{\bf b}\,-\,{\bf L}({\bf a}\times{\bf b},\,\lambda=-1) \notag \\
&=-\,{\bf a}\cdot{\bf b}\,+\,{\bf D}({\bf a}\times{\bf b}) \notag \\
&=-\,{\bf b}\cdot{\bf a}\,-\,{\bf D}({\bf b}\times{\bf a}) \notag \\
&={\bf D}({\bf b})\;{\bf D}({\bf a}). 
\end{align}
It is now evident from the above equations that, once the relevant products are correctly evaluated, the ordering relation between the spin bivectors ${{\bf L}({\bf a},\,\lambda)}$ and ${{\bf L}({\bf b},\,\lambda)}$ and the detector bivectors ${{\bf D}({\bf a}})$ and ${{\bf D}({\bf b})}$ is equivalent to the hypothesis that the orientation $\lambda=\pm1$ of the 3-sphere is a fair coin. This disproves the conjecture in the critique \cite{Gill-Ieee} that equations (34) and (35) of \cite{IEEE-2} could not both be correct.

The rest of the argument in the critique pertaining to the validity of the identity (\ref{50a}) is now moot. But it is instructive to unpack it to appreciate the mistake in it. In fact, it is an elementary mistake in ordinary vector algebra, as pointed out in Appendix~C of \cite{Reply-Gill}. The critique considers both right-handed and left-handed cross products. But there is no such suggestion in \cite{Disproof,IJTP,RSOS,IEEE-1,IEEE-2}. Contrary to what the critique considers, namely, two cross products $\times_{\!\lambda}$ with $\lambda =  \pm 1$ and the rules 
\begin{equation}
\mathbf  a \times_{\!+1}\! \mathbf b ~=~ \mathbf a \times \mathbf b \qquad \text{and} \qquad \mathbf a \times_{\!-1}\! \mathbf b ~=~ \mathbf b \times \mathbf a, \label{inc}
\end{equation}
the correct versions of these equations in vector algebra are
\begin{equation}
\mathbf  a \times_{\!+1}\! \mathbf b ~=~ \mathbf a \times \mathbf b \qquad \text{and} \qquad \mathbf a \times_{\!-1}\! \mathbf b ~=~ \mathbf a \times \mathbf b.
\end{equation}
As pointed out in Appendix~C of \cite{Reply-Gill}, in vector algebra the cross product between vectors ${\bf a}$ and ${\bf b}$ remains the same in both right-handed and left-handed coordinates. But by using the incorrect equations (\ref{inc}), the critique arrives at its Eq.~(7),
\begin{equation}
\mathbf L(\mathbf a, \lambda) \mathbf L(\mathbf b, \lambda) ~=~ - \mathbf a \cdot \mathbf b - \mathbf L(\mathbf a \times_{\!\lambda} \!\mathbf b, \lambda),
\end{equation}
which, according to its definition of $\mathbf a \times_{\!\lambda} \!\mathbf b$, is equivalent to 
\begin{equation}
\mathbf L(\mathbf a,\,\lambda) \mathbf L(\mathbf b,\,\lambda) ~=~ - \mathbf a \cdot \mathbf b - \,\lambda\;\mathbf L(\mathbf a \times \mathbf b,\, \lambda).
\end{equation}
Note that there is now a redundant $\lambda$ on the right-hand side of this equation in the second term, in contrast to the identity (\ref{50a}) above. In terms of the basis, the equation takes the form 
\begin{equation}
L_i(\lambda)\,L_j(\lambda) = -\,\delta_{ij} -\lambda \sum_k\epsilon_{ijk}\,L_k(\lambda),
\end{equation}
with the extra $\lambda$ appearing just before the summation. The critique now claims that, with this equation, it has ``restored consistency.'' But, in fact, what it has done is to introduce {\it inconsistency}. For the above equation {\it does not} represent the bivector subalgebra of ${\mathrm{Cl}}_{(3,0)}$, and therefore it cannot constitute a quaternionic 3-sphere. The correct bivector subalgebra of ${\mathrm{Cl}}_{(3,0)}$ is given by (\ref{wh-o8899}) discussed above, without the extra $\lambda$. Thus, unlike Eq.~(28) of \cite{IEEE-2}, the critique's Eq.~(7) with the extra $\lambda$ is both conceptually and mathematically incorrect.

\subsection{Concerning Equations (39) and (40) of \cite{IEEE-2}}

\begin{figure}[t]
\centering
\includegraphics[scale=0.8]{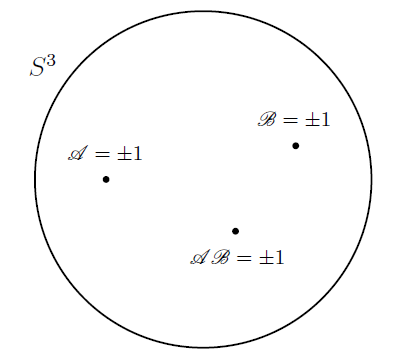}
\caption{The results ${\mathscr A}$ and ${\mathscr B}$ are simultaneous points of a quaternionic 3-sphere, or ${S^3}$. Since ${S^3}$ remains closed under multiplication, the product ${{\mathscr A}{\mathscr B}}$ is also a point of ${S^3}$, with its binary value ${\pm1}$ dictated by the geometry.}
\label{Fig-1}
\end{figure}

The critique's second incorrect claim concerns the definitions (39) and (40) in \cite{IEEE-2} of measurement functions, which I have reproduced in (\ref{53}) and (\ref{54}) above. The critique claims that the functions (39) and (40) predict  perfect anti-correlation 
\begin{equation}
{\cal E}({\bf a},\,{\bf b}) = -1 
\end{equation}
for all choices of measurement directions ${\mathbf a}$ and ${\mathbf b}$, instead of the cosine correlations derived in (\ref{65a}) for ${{\mathbf a}\not={\mathbf b}}$. This claim is refuted in Appendix~C of \cite{Reply-Gill} and in \cite{Scott}, as well as in Answers~9 and 14 in Appendix~B of \cite{IEEE-1} and Answers~6 and 7 in Appendix~B of \cite{IEEE-2}. But the claim is made again in this critique, in Eq.~(3) near the end of its Section II, and repeated in the paragraph that includes Eqs.~(9) and (10), which read
\begin{align}
{\mathscr A}({\bf a},\lambda) &= +\lambda, \label{idA} \\
{\mathscr B}({\bf b},\lambda) &= -\lambda. \label{idB}
\end{align}
In other words, the critique {\it identifies} the measurement results ${\mathscr A}$ and ${\mathscr B}$ observed at the space-like separated stations with the initial state $\lambda$ of the spins that originates from the source located in the overlap of the backward light-cones of Alice and Bob. But no such identification is made in the definitions (39) and (40) of the measurement functions in \cite{IEEE-2}. What is encapsulated by these functions are the measurement interactions. Alice chooses her detector ${\bf D}({\bf a})$ about a measurement direction ${\bf a}$ and Bob chooses his detector ${\bf D}({\bf b})$ about a measurement direction ${\bf b}$, at a space-like distance from each other. The spins ${-\,{\bf L}({\bf s}_1,\,\lambda)}$ and ${+\,{\bf L}({\bf s}_2,\,\lambda)}$, on the other hand, originate from the source located in the overlap of the backward light-cones of Alice and Bob. And the results ${\mathscr A}({\bf a},\lambda)$ and  ${\mathscr B}({\bf b},\lambda)$ defined in (39) and (40) do not come about until the time of measurements, and even then only via two different quaternions within $S^3$. The identifications (\ref{idA}) and (\ref{idB}), however, allow the critique to write their product as
\begin{equation}
{\mathscr A}({\bf a},\lambda)\,{\mathscr B}({\bf b},\lambda) = (+\lambda)(-\lambda) = -\lambda^2 = -1,
\end{equation}
for all choices of ${\bf a}$ and ${\bf b}$. But for ${{\bf a}\not={\bf b}}$, Eqs.~(\ref{idA}) and (\ref{idB}) are valid only for $\mathbf s_1 \not= \mathbf s_2$. They ignore the conservation of zero spin angular momentum, which, as we saw in Eq.~(\ref{56a}), amounts to setting ${\mathbf s}_1={\mathbf s}_2={\bf s}$. In other word, the critique's equations (9) and (10) hold in general for all choices of ${\bf a}$ and ${\bf b}$ {\it if and only if} the conservation of spin angular momentum is violated, or, equivalently, the M\"obius-like twists in the Hopf bundle of $S^3$ are ignored. That is to say, for ${{\bf a}\not={\bf b}}$, equations (\ref{idA}) and (\ref{idB}) hold if and only if the 3-sphere model is abandoned and one stoops back to the flat geometry of ${\mathrm{I\!R}^3}$.

For convenience, I have reproduced one of the correct derivations of the singlet correlations above in the paragraphs containing equations (\ref{16a}) to (\ref{65a}). This derivation preserves the geometrical properties of the 3-sphere, without relapsing back to the flat geometry of ${\mathrm{I\!R}^3}$ as the critique tends to do. It is also worth noting that the derivation in (\ref{16a}) to (\ref{65a}) above is just one of several different ways it is demonstrated in \cite{IEEE-1} and \cite{IEEE-2} that within a quaternionic 3-sphere, taken as a physical space, the correlations are inevitably ${\cal E}({\bf a},\,{\bf b}) = -{\bf a}\cdot{\bf b}$.

There is also a related conceptual issue that is important to address here. Recall that there are, in fact, three different sets of experiments involved in any EPR-Bohm type experiments. Alice and Bob can independently detect spins of the particles they receive at their respective observation stations, obtaining the results $\pm1$, with 50/50 chance, so that both $\langle {\mathscr A} \rangle = 0$ and $\langle {\mathscr B} \rangle = 0$. These are two separate sets of experiments, because Alice can perform her experiments and obtain the same results regardless of Bob’s existence, and vice versa. Their separate and independent results, ${\mathscr A} = \pm1$, $\langle {\mathscr A} \rangle = 0$ and ${\mathscr B} = \pm1$, $\langle {\mathscr B} \rangle = 0$, respectively, are exactly what the measurement functions (39) and (40) defined in \cite{IEEE-2} predict.

Then, in a third set of experiments, Alice and Bob {\it jointly} and {\it simultaneously}, but again independently, detect spins at their respective stations, regardless of which spin result the other party has observed. Their recorded results are then compared later by a third party, say Charlie, and calculated to exhibit the correlation $\langle{\mathscr A}{\mathscr B} = \pm1\rangle = -\cos(\,\eta_{{\bf a}{\bf b}})$ between the results ${\mathscr A} = \pm1$ and ${\mathscr B} = \pm1$. In actual experiments this can be done {\it only} by ``coincidence counts'' of joint and simultaneous detections of spins by Alice and Bob (cf. Section 4.1 of \cite{RSOS}). It is therefore a mistake to confuse the first two sets of experiments with the third set of experiments.

Now recall that the results $\mathscr{A}$ and $\mathscr{B}$ observed simultaneously but independently by Alice and Bob are necessarily events in spacetime. Within the spacetime defined by the line element (\ref{frw}) together with $\Sigma=S^3$ and $a(t)=1$, they are thus events in ${\mathrm{I\!R}\times S^3}$. Thus, recalling that $S^3$ --- which, as defined in (\ref{nonsin}), is a set of unit quaternions --- remains closed under multiplication (cf. Fig.~\ref{Fig-1}), the correct question to ask here is the one I have posed just before Eq.~(\ref{16a}) in Section \ref{I}: What will be the average of the product ${{\mathscr A}{\mathscr B}}$ of the results ${\mathscr A}$ and ${\mathscr B}$ within the space-like hypersurface $S^3$? In other words, what will be the average of the product ${{\mathscr A}{\mathscr B}}$ when the results ${\mathscr A}$ and ${\mathscr B}$ are observed by Alice and Bob separately but simultaneously, in ``coincidence counts'', within $S^3$? We can only work out the correct average of ${{\mathscr A}{\mathscr B}}$ within $S^3$ from the quaternionic definitions (39) and (40), not from their travesties (\ref{idA}) and (\ref{idB}), and that answer, contrary to the claim in \cite{Gill-Ieee}, works out to be ${\cal E}({\bf a},\,{\bf b}) = -{\bf a}\cdot{\bf b}$.

\subsection{Concerning a pair of binary variables}

In the paragraph before last in its Section II, the critique \cite{Gill-Ieee} claims:
\begin{quote}
For $\mathbf a \ne \pm \mathbf b$, the probability distribution of the pair of binary variables $(X_{\mathbf a}, Y_{\mathbf b})$ predicted by quantum mechanics gives positive probability to each of four distinct joint outcomes $(\pm1, \pm1)$. There is no way one can simulate a single draw from a probability distribution over four outcomes, each  of positive probability, as a deterministic function of the outcome of {\it one} fair coin toss. Christian's hidden variable $\lambda$, which one may identify with the elementary outcome $\omega$ of the alleged probability model on which all those random variables are defined, is a fair coin toss, and in his model, the results of measurement of spin of the two particles in any two directions are functions only of $\lambda$ and of the relevant direction.
\end{quote}
This argument is again based on a mistaken reading of what the hidden variable $\lambda$ is in the $S^3$ model. In the $S^3$ model ${\lambda}$ is not a detached fair coin toss in the flat space ${\mathrm{I\!R}^3}$. Instead, it represents an orientation or handedness of the closed and compact space $S^3$ itself. The alternatives are thus between an inside-out 3-sphere and outside-out\footnotemark[1] 3-sphere, with respect to the detectors ${\bf D}({\bf a})$ and ${\bf D}({\bf b})$. This again illustrates that the critique has misinterpreted what the 3-sphere model actually is and how it predicts the correlations ${\cal E}({\bf a},\,{\bf b}) = -{\bf a}\cdot{\bf b}$.

\subsection{Concerning the computer Code in \cite{IEEE-2}}

Next, the critique \cite{Gill-Ieee} turns to the computer code presented in \cite{IEEE-2} and claims that it contains the following ``revealing line'':
\begin{equation}
\texttt{if(lambda==1) \{q=A B;\} else \{q=B A;\}}, \notag
\end{equation}
``which serves to switch between the geometric product and its transpose according to the sign of $\lambda$'' \cite{Gill-Ieee}. But it is easy to see that the above line is a faithful representation of the analytical equations (34) and (35) in \cite{IEEE-2} that switch the order of the detectors with respect to that of the spins (as annotated in the code), thereby shuffling the alternative orientations of $S^3$, exactly as required by those equations. In other words, the line in question is necessitated by the model itself. In fact, the line in question is {\it the very essence} of the 3-sphere model. Moreover, by now the code has been independently translated by several professional programmers into different computer languages, such as {\it Python}, {\it Maple}, {\it R}, and {\it Mathematica} \cite{Fred}.

It is worth stressing here that the computer code included in \cite{IEEE-2}, by itself, is {\it not} the model, or even a proof of the model. Its purpose is to demonstrate how the model works. It is thus a pedagogical tool that verifies the analytical computations presented in \cite{IEEE-2}. Needless to say, the analytical computations stand on their own and do not require a numerical simulation for their validity. On the other hand, the computer code for an event-by-event simulation of the singlet correlations does provide additional support to the analytical computations, for it is both pedagogically and statistically illuminating. 

By contrast, in the paragraph before last in its Section~II, and then at the end of its Section~III, the critique~\cite{Gill-Ieee} claims:
\begin{quote}
Christian's computer simulation program uses a fair coin toss to average Geometric Algebra products using the fundamental GA formula $\mathbf a\cdot \mathbf b =\frac12 \mathbf a \mathbf b + \frac12 \mathbf b \mathbf a$. ... About half the time, the ``product of the measurements'' is {\it defined by the code} as the quaternion $-\mathbf a \mathbf b$, the other half of the time it is the quaternion $- \mathbf b \mathbf a$.
\end{quote}
But these claims are not correct. According to the correct algorithm, which is annotated in the code, the orientation $\lambda$ of the 3-sphere plays the role of a fair coin quite independently of the definition of the geometric product. Moreover, contrary to the claim in \cite{Gill-Ieee}, the values of the scalar part of the product ${\bf a}{\bf b}$ are not ``grouped into bins.'' Instead, what is being plotted is the scalar part of the correlation directly with respect to its associated angle $\eta_{{\bf a}{\bf b}}$. The purpose of the code is to verify the analytical calculations, which it does quite successfully.

\section{Concluding remarks}

After summarizing the local-realistic 3-sphere model for the singlet correlations presented in \cite{IEEE-2}, I have demonstrated that none of the claims made against the model in the critique \cite{Gill-Ieee} are correct. In particular, contrary to its claim, the critique has not identified any mistakes in \cite{IEEE-2}, either in the analytical model of the singlet correlations or in its event-by-event numerical simulation. The critique begins with an incorrect version of the 3-sphere model by writing some of its equations incorrectly. It then derives a constant value for the singlet correlations by failing to respect the geometrical properties of the 3-sphere such as the spinorial sign changes in the quaternions, criticizes this incorrect value, and concludes that it has thereby criticized the 3-sphere model. This strategy also violates the conservation of spin angular momentum. In this paper I have addressed the claims in the critique \cite{Gill-Ieee} point by point and shown that they are neither proven nor justified.

\end{document}